\documentclass[12pt,hyper,notoc]{JHEP3}
\usepackage{graphics,psfrag}
\usepackage{amsmath,amsthm,amssymb,epsfig,euscript,array,cite,cancel}
\usepackage{cases,empheq}
\usepackage{verbatim}

\theoremstyle{definition}

\theoremstyle{remark}

\newcounter{multieqs}

\newcommand{\be}{\begin{equation}}
\newcommand{\ee}{\end{equation}}
\newcommand{\eq}[1]{(\ref{#1})}
\newcommand{\bit}{\begin{itemize}}  \newcommand{\eit}{\end{itemize}}

\newcommand{\bm}[1]{\mbox{\boldmath $#1$}}
\newcommand{\rf}[1]{(\ref{#1})}

\def\bd{\begin{document}}
\def\ed{\end{document}}
\def\nn{\nonumber}
\def\bea{\begin{eqnarray}}
\def\eea{\end{eqnarray}}
\let\bm=\bibitem

\def\la{\langle}
\def\ra{\rangle}

\def\npb#1#2#3{Nucl. Phys. {\bf{B#1}} #3 (#2)}
\def\plb#1#2#3{Phys. Lett. {\bf{#1B}} #3 (#2)}
\def\prl#1#2#3{Phys. Rev. Lett. {\bf{#1}} #3 (#2)}
\def\prd#1#2#3{Phys. Rev. {D \bf{#1}} #3 (#2)}
\def\cmp#1#2#3{Comm. Math. Phys. {\bf{#1}} #3 (#2)}
\def\cqg#1#2#3{Class. Quantum Grav. {\bf{#1}} #3 (#2)}
\def\nppsa#1#2#3{Nucl. Phys. B (Proc. Suppl.) {\bf{#1A}}#3 (#2)}
\def\ap#1#2#3{Ann. of Phys. {\bf{#1}} #3 (#2)}
\def\ijmp#1#2#3{Int. J. Mod. Phys. {\bf{A#1}} #3 (#2)}
\def\rmp#1#2#3{Rev. Mod. Phys. {\bf{#1}} #3 (#2)}
\def\mpla#1#2#3{Mod. Phys. Lett. {\bf A#1} #3 (#2)}
\def\jhep#1#2#3{J. High Energy Phys. {\bf #1} #3 (#2)}
\def\atmp#1#2#3{Adv. Theor. Math. Phys. {\bf #1} #3 (#2)}

\def\N{{\cal N}}
\def\sst{\scriptscriptstyle}
\def\thetabar{\bar\theta}
\def\Tr{{\rm Tr}}
\def\one{\mbox{1 \kern-.59em {\rm l}}}

\def\a{\alpha}      \def\da{{\dot\alpha}}  \def\dA{{\dot A}}
\def\b{\beta}       \def\db{{\dot\beta}}  
\def\g{\gamma}  \def\G{\Gamma}  \def\dc{{\dot\gamma}}  
\def\d{\delta}  \def\D{\Delta}  \def\ddt{\dot\delta}  
\def\e{\epsilon}        \def\ve{\varepsilon}  
\def\f{\phi}    \def\F{\Phi}    \def\vvf{\f}  
\def\h{\eta}  
\def\k{\kappa}  
\def\l{\lambda} \def\L{\Lambda}  
\def\m{\mu} \def\n{\nu}  
\def\o{\omega}  
\def\p{\pi} \def\P{\Pi}  
\def\r{\rho}  
\def\s{\sigma}  \def\S{\Sigma}  
\def\t{\tau}  
\def\th{\theta} \def\Th{\Theta} \def\vth{\vartheta}  
\def\X{\Xeta}  
\def\z{\zeta}  

\def\na{\nabla}  

\def\cA{{\cal A}} \def\cB{{\cal B}} \def\cC{{\cal C}}  
\def\cD{{\cal D}} \def\cE{{\cal E}} \def\cF{{\cal F}}  
\def\cG{{\cal G}} \def\cH{{\cal H}} \def\cI{{\cal I}}  
\def\cJ{{\cal J}} \def\cK{{\cal K}} \def\cL{{\cal L}}  
\def\cM{{\cal M}} \def\cN{{\cal N}} \def\cO{{\cal O}}  
\def\cP{{\cal P}} \def\cQ{{\cal Q}} \def\cR{{\cal R}}  
\def\cS{{\cal S}} \def\cT{{\cal T}} \def\cU{{\cal U}}  
\def\cV{{\cal V}} \def\cW{{\cal W}} \def\cX{{\cal X}}  
\def\cY{{\cal Y}} \def\cZ{{\cal Z}}

\def\ua{\underline{\alpha}}  
\def\uc{\underline{\phantom{\alpha}}\!\!\!\gamma}  
\def\um{\underline{\mu}}  
\def\ud{\underline\delta}  
\def\ue{\underline\epsilon}  
\def\una{\underline a}\def\unA{\underline A}  
\def\unb{\underline b}\def\unB{\underline B}  
\def\unc{\underline c}\def\unC{\underline C}  
\def\und{\underline d}\def\unD{\underline D}  
\def\une{\underline e}\def\unE{\underline E}  
\def\unf{\underline{\phantom{e}}\!\!\!\! f}\def\unF{\underline F}  
\def\unm{\underline m}\def\unM{\underline M}  
\def\unn{\underline n}\def\unN{\underline N}  
\def\unp{\underline{\phantom{a}}\!\!\! p}\def\unP{\underline P}  
\def\unq{\underline{\phantom{a}}\!\!\! q}  
\def\unQ{\underline{\phantom{A}}\!\!\!\! Q}  
\def\unH{\underline{H}}  
  
\def\As {{A \hspace{-6.4pt} \slash}\;}  
\def\bs {{b \hspace{-6.4pt} \slash}\;}  
\def\Ds {{D \hspace{-6.4pt} \slash}\;}
\def\Gts {{\Gt \hspace{-6.4pt} \slash}\;}
\def\ds {{\del \hspace{-6.4pt} \slash}\;}  
\def\ss {{\s \hspace{-6.4pt} \slash}\;}  
\def\ks {{ k \hspace{-6.4pt} \slash}\;}  
\def\ps {{p \hspace{-6.4pt} \slash}\;}   
\def\xs {{x \hspace{-6.4pt} \slash}\;}  
\def\pas {{{p_1} \hspace{-6.4pt} \slash}\;}  
\def\pbs {{{p_2} \hspace{-6.4pt} \slash}\;}   
\def\cFs {{{\cal F} \hspace{-6.4pt} \slash}\;}

\def\Ah{{\hat{A}}}  
\def\Dh{{\hat{D}}}
\def\Gh{{\hat{G}}}
\def\Fh{{\hat{F}}}
\def\Ih{{\hat{I}}} 
\def\Jh{{\hat{J}}} 
\def\Kh{{\hat{K}}}
\def\Lh{{\hat{L}}} 
\def\Ph{{\hat{P}}}
\def\Rh{{\hat{R}}}
\def\Vh{{\hat{V}}} 
\def\Xh{{\hat{X}}}
 
\def\ah{{\hat{\a}}}
\def\bh{{\hat{\b}}}
\def\gh{{\hat{\g}}}
\def\dh{{\hat{\d}}}
\def\hh{\hat{h}}
\def\uh{\hat{u}}  
\def\xh{\hat{x}}  
\def\yh{\hat{y}}  
\def\ph{\hat{p}}  
\def\xih{\hat{\xi}}  
\def\chih{\hat{\chi}}  
\def\Psih{\hat{\Psi}}    
\def\phih{\hat{\phi}}

\def\psit{\tilde{\psi}}  
\def\Psit{\tilde{\Psi}}   
\def\Psibt{\tilde{\bar{Psi}}}  

\def\st{\tilde{\sigma}}  

\def\delt{\tilde{\delta}}
\def\Phit{\tilde{\Phi}}   
\def\Phitb{\overline{\tilde{Phi}}}  
\def\tht{\tilde{\th}}  
\def\lt{\tilde{\l}}
\def\chit{\tilde{\chi}}   
\def\phit{\tilde{\phi}} 

\def\At{\tilde{A}}
\def\Bt{\tilde{B}}
\def\Ct{\tilde{C}}
\def\Dt{\tilde{D}}
\def\Et{\tilde{E}}
\def\Ft{\tilde{F}}
\def\Gt{\tilde{G}}
\def\Ht{\tilde{H}}
\def\It{\tilde{I}}
\def\Jt{\tilde{J}}
\def\Qt{\tilde{Q}}  
\def\Rt{\tilde{R}}  
\def\Mt{\tilde{M }}  
\def\Nt{\tilde{N}}   
\def\St{\tilde{S}}
\def\Vt{\tilde{V}}
\def\Xt{\tilde{X}} 
\def\at{\tilde{a}}
\def\ct{\tilde{c}}
\def\dt{\tilde{d}}
\def\htt{\tilde{h}} 
\def\ft{\tilde{f}}
\def\gt{\tilde{g}}
\def\pt{\tilde{p}}  
\def\qt{\tilde{q}}  
\def\vt{\tilde{v}}  
\def\nt{\tilde{n}}  
\def\ut{\tilde{u}}  
\def\wt{\tilde{w}}  
\def\zt{\tilde{z}} 
\def\xt{\tilde{x}} 
\def\yt{\tilde{y}} 
\def\Psit{\tilde{\Psi}}
\def\vphit{\tilde{\varphi}}  

\def\eb{\bar{\epsilon}} 
\def\delb{\bar{\partial}}  
\def\thb{\bar{\theta}}
\def\mub{\bar{\mu}}
\def\lamb{\bar{\l}}
\def\psib{\bar{\psi}}
\def\sb{\bar{\sigma}}
\def\xib{\bar{\xi}}
\def\chib{\bar{\chi}}

\def\Psib{\bar{\Psi}}
\def\Phib{\bar{\Phi}}
\def\Lamb{\bar{\Lambda}}
\def\Sb{{\overline \Sigma}}
\def\cb{\bar{c}}
\def\hb{\bar{h}}
\def\qb{\bar{q}}
\def\wb{\bar{w}}
\def\ub{\bar{u}}
\def\zb{{\bar{z}}}
\def\Hb{\bar{H}}
\def\Qb{{\bar Q}}
\def\Omegab{\overline{\Omega}}
\def\ob{\overline{\omega}}

\def\Ab{{\overline A}} \def\Bb{{\overline B}} \def\Cb{{\overline C}}  
\def\Db{{\overline D}} \def\Eb{{\overline E}} \def\Fb{{\overline F}}  
\def\Gb{{\overline G}} 
\def\Ib{{\overline I}}  
\def\Jb{{\overline J}} \def\Kb{{\overline K}} \def\Lb{{\overline L}}  
\def\Mb{{\overline M}} \def\Nb{{\overline N}} \def\Ob{{\overline O}}  
\def\Pb{{\overline P}}  \def\Rb{{\overline R}}  
 \def\Tb{{\overline T}} \def\Ub{{\overline U}}  
\def\Vb{{\overline V}} \def\Wb{{\overline W}} \def\Xb{{\overline X}}  
\def\Yb{{\overline Y}} \def\Zb{{\overline Z}}  

\def\fb{{\overline f}}
\def\gb{{\overline g}}
\def\mb{{\overline m}}
\def\lb{{\overline l}}
\def\yb{{\overline y}}
  
\def\ldel{{\overleftarrow{\del}}}
\def\rdel{{\overrightarrow{\del}}}
\def\ldeldel{{\overleftarrow{\del^2}}}
\def\rdeldel{{\overrightarrow{\del^2}}}
\def\ldelb{{\overleftarrow{\bar{\del}}}}
\def\rdelb{{\overrightarrow{\bar{\del}}}}

\def\ba{{\bf a}} 
\def\bk{{\bf k}}  
\def\bl{{\bf l}}  
\def\bp{{\bf p}}  
\def\bq{{\bf q}}  
\def\br{{\bf r}}
\def\bt{{\bf t}}
\def\bu{{\bf u}}
\def\bv{{\bf v}}
\def\bx{{\bf x}}  
\def\by{{\bf y}}  
\def\bR{{\bf R}}  
\def\bV{{\bf V}}

\def\bone{{\bf 1}}  

\def\va{{\vec a}}
\def\vk{{\vec k}}
\def\vp{{\vec p}}
\def\vq{{\vec q}}
\def\vx{{\vec x}}
\def\vy{{\vec y}}
\def\vu{{\vec u}}
\def\vv{{\vec v}}
\def \vH{{\vec H}}
\def \vg{{\vec g}}

\def\vs{{\vec \sigma}}
\def\vtau{{\vec \tau}}

\newcommand{\ov}[1]{\overrightarrow{#1}}

\def\frA{\mathfrak{A}}
\def\frB{\mathfrak{B}}
\def\frC{\mathfrak{C}}
\def\frD{\mathfrak{D}}
\def\frE{\mathfrak{E}}
\def\frF{\mathfrak{F}}
\def\frG{\mathfrak{G}}
\def\frH{\mathfrak{H}}
\def\frM{\mathfrak{M}}
\def\frN{\mathfrak{N}}
\def\frR{\mathfrak{R}}
\def\frW{\mathfrak{W}}

\def\fra{\mathfrak{a}}
\def\frb{\mathfrak{b}}
\def\frf{\mathfrak{f}}
\def\frg{\mathfrak{g}}
\def\frh{\mathfrak{h}}
\def\frl{\mathfrak{l}}
\def\frs{\mathfrak{s}}
\def\fri{\mathfrak{i}}
\def\frj{\mathfrak{j}}

\def\ma{\mathfrak{a}}
\def\mg{\mathfrak{g}}
\def\mh{\mathfrak{h}}
\def\mR{\mathfrak{R}}
\def\mN{\mathfrak{N}}

\def\d{\delta}\def\D{\Delta}\def\ddt{\dot\delta}  
  
\def\pa{\partial} \def\del{\partial}  
\def\xx{\times}  
\def\uno{\mbox{1 \kern-.59em {\rm l}}}    
  
\def\trp{^{\top}}  
\def\inv{^{-1}}  
\def\dag{{^{\dagger}}}  
\def\pr{^{\prime}}  
  
\def\rar{\rightarrow}  
\def\lar{\leftarrow}  
\def\lrar{\leftrightarrow}  
  
\newcommand{\0}{\,\!}      
\def\one{1\!\!1\,\,}  
\def\im{\imath}  
\def\jm{\jmath}  
  
\newcommand{\tr}{\mbox{tr}}  
\newcommand{\slsh}[1]{/ \!\!\!\! #1}  
  
\def\vac{|0\rangle}  
\def\lvac{\langle 0|}  
  
\def\hlf{\frac{1}{2}}  
\def\ove#1{\frac{1}{#1}}  

\def\Box{\square}  
\def\CC {\mathbb{C}}
\def\FF {\mathbb{F}}
\def\RR{\mathbb{R}}
\def\NN{\mathbb{N}}  
\def\ZZ{\mathbb{Z}}  
\def\bb#1{{\bf #1}}  
\def\bcomment#1{}  
\def\bfhat#1{{\bf \hat{#1}}}  
\def\VEV#1{\left\langle #1\right\rangle}  

\newcommand{\ex}[1]{{\rm e}^{#1}} \def\ii{{\rm i}}  

\newcommand{\lrbrk}[1]{\left(#1\right)}
\newcommand{\lrsbrk}[1]{\left[#1\right]}
\newcommand{\sfrac}[2]{{\textstyle\frac{#1}{#2}}}
 
\def\stw{{\sqrt{2}}}

\def\rf {{\rm f}}
\def\ri {{\rm i}}
\def\rj {{\rm j}}
\def\rk {{\rm k}}
\def\rl {{\rm l}}
\def\rs {{\scriptscriptstyle \rm S}}
\def\rt {{\scriptscriptstyle \rm T}}

\def\rQ {{\scriptscriptstyle \rm \cQ}}
\def\rR {{\scriptscriptstyle \rm \cR}}

\def\cQb{{\cal \Qb}}
\def\cRb{{\cal \Rb}}
\def\cWb{{\cal \Wb}}

\def\fd {{\rm N}}
\def\afd {{\overline{\rm N}}}

\def \II {I\hspace{-.1em}I\hspace{.1em}}
\def \IIA {\mbox{\II A\hspace{.2em}}}
\def \IIB {\mbox{\II B\hspace{.2em}}}
\def \gs {g^s}
\def \ls {\lambda^s}

\def \I {{\cal I}}
\def \qs {q\hspace{-.53em}/\hspace{.15em}}
\def \ks {k\hspace{-.53em}/\hspace{.15em}}
\def \YM {{\mbox{\tiny YM}}}
\def \gym {g_{\YM}}

\def \Lc {\L_c}
\def\IR{\relax{\rm I\kern-.18em R}}
\def \id {{\bf 1}}

\def\cci{\ell}
\def\ccj{\ell'}

\def \thbb{\overline{\th\th}}
\newcommand \ol{\overline}
\def \lamb{\bar{\lambda}}
\def \vphi{\varphi}
\def \lambh{\hat{\bar{\lambda}}}
\def \lh{\hat{\lambda}}
\def \dd{\ddagger}

\newcommand{\QNB}[3]{[#1,#2,#3]}
\def\hm{\tilde{\eta}} 
\def\lp{l_{+}}
\def\lm{l_{-}}
\def \PS {{(\text{PS})}}
\def \Dir {{(\text{Dirac})}}
\def \WY {{(\text{WY})}}
\def \Sin {{(\text{Sin})}}
\def \tHP{{(\text{'t-P})}}
\def \uo {{U(1)}}
\def \Lt {\tilde{L}}
\def \tn {{\tau}^{(n)}}
\def \rn {{\hat{r}}^{(n)}}
\def \thn {{\hat{\theta}}^{(n)}}
\def \vphin {{\hat{\vphi}}^{(n)}}

\author{Chong-Sun Chu$^{1,2}$ 
and Pichet Vanichchapongjaroen$^2$ \\
$^1$ Department of Physics and
National Center for Theoretical Sciences \\
National Tsing-Hua University, Hsinchu 30013, Taiwan \\
$^2$ Centre for Particle Theory and Department of Mathematics, \\
Durham University, Durham, DH1 3LE, UK\\
E-mail:  
\email{chong-sun.chu@durham.ac.uk},
\email{pichet.vanichchapongjaroen@durham.ac.uk}
}

\title{
Non-abelian Self-Dual String and \\M2-M5 Branes Intersection in Supergravity
}

\abstract{
We consider the non-abelian theory
\href{http://arxiv.org/abs/arXiv:1203.4224}{\cite{CK}} 
for an arbitrary number $N_5$
of five-branes and construct
self-dual string solution with an arbitrary  $N_2$ unit of self-dual
charges. This generalizes the previous solution of non-abelian 
self-dual string \href{http://arxiv.org/abs/arXiv:1207.1095}{ \cite{CKV}}
of $N_5=2, N_2=1$.
The 
radius-transverse distance relation describing the M2-branes spike,
particularly its dependence on $N_2$ and $N_5$, is obtained
and is found to agree precisely
with the supergravity description of an intersecting M2-M5 branes 
system. 
} 

\preprint{DCPT-13/13}

\keywords{M-Theory, D-branes, M-branes, Gauge Symmetry}

\begin{document}

\section{Introduction}

The low energy theory of $N$ coincident M5-branes in a flat spacetime 
is given by an interacting (2,0) superconformal
theory in six dimensions. The understanding of the
dynamics of this system is of utmost importance. There exists a number of 
proposals
for the fundamental definition of the quantum theory. For example, in terms of 
DLCQ instanton quantum mechanics \cite{qm1,qm2}, deconstruction \cite{dec}, 
and more recently in terms of the 5d supersymmetric Yang-Mills theory
\cite{dou,lam2}.  
The 5-dimensional maximally supersymmetric Yang-Mills theory 
has been shown to be consistent with what we know about M5-branes   
\cite{ta} - \cite{kkk}, as well as on its partition function 
\cite{lee} - \cite{bg}.
Despite the appearance of a UV divergence \cite{bern}, 
it is possible  that the theory is non-perturbatively well-defined
\footnote{It is important to understand how this happens, which will 
deepen our understanding about quantum field theory. It would 
also
have implications
on the renormalizability of gravity theory.}
.

Recently, a six-dimensional action principle for the low energy theory has  been
proposed \cite{CK}.
This construction  was motivated by the 
analysis in \cite{chu,CS} where  
a set of 5d Yang-Mills gauge fields was introduced in order to
incorporate non-trivial interactions among the 2-form potential.  
An important feature of this theory
is that the self-interaction of the two-form gauge field is mediated
by a  Yang-Mills gauge field which  
is constrained to be determined completely in terms of the non-abelian
tensor gauge field 
\be\label{FB1}
F_{\m\n} = c\int dx_5 \Ht_{\m\n}, 
\ee
and hence is auxiliary. This is necessary since there is no room in 
the (2,0) tensor
multiplet to accommodate additional propagating degrees of freedom.
 
Although one does not have the full supersymmetric construction of the 
non-abelian (2,0) theory, by combining knowledge of conformal symmetry and 
R-symmetry one can argue for the form of the 1/2 BPS equations in the 
case when only one scalar field is turned on \cite{CKV}.
In the follow up paper \cite{CKV},  the case of two M5-branes was considered
and the $SU(2)$ non-abelian 5-branes equation was solved. 
It was argued that the solution 
could be lifted to become  a solution of the non-abelian (2,0) theory
with  self-dual electric and magnetic charges.
This is precisely 
the relation between the pure 
self-dual string solution in the 
linear Perry-Schwarz \cite{PS} 
and the half BPS solution of Howe-Lambert-West \cite{HLW}.
It was also shown that the dimensionless 
constant $c$, which is a free parameter in the action, 
is fixed by  the charge quantization of the self-dual string solution 
in the theory. As a result, the  solution carries a unit of
self-dual charges and  
describes  an M2-brane spike emerging out of the 
multiple M5-branes worldvolume, 
with the adjoint scalar representing the transverse dimension.

In the next section, 
we generalize this construction to 
the general case of an arbitrary number $N_5$ of 
M5-branes and construct a self-dual string solution with $N_2$ unit of self-dual
charges. Like the unit charge solution in \cite{CKV}, 
the new self-dual string obtains its charge from the auxiliary Yang-Mills 
field configuration,
which is now given by a generalization
of the $SU(2)$ Wu-Yang monopole solution to one with arbitrary charge $N_2$. 
We will show that the value of $c$ is again fixed by the requirement of 
charge quantization. 
Moreover we will work out the dependence on $N_2$ and $N_5$ in the 
radius-transverse distance relation describing the M2-branes spike, and show that
it agrees with the supergravity description of an intersecting M2-M5 branes 
system. Therefore our results provide further support of the model of \cite{CK}.
The paper is concluded with some further discussions in section 3. 

Other related works on multiple M5-branes include: \cite{lam1} constructed
a non-abelian version of (2,0)
supersymmetric equation of motion using Lie 3-algebra;
twistor space formulation of M5-branes has been proposed 
\cite{twistor1,twistor2,twistor3};
\cite{ho,bon} constructed a compactified theory of 
non-abelian 2-form gauge potentials with a self-dual field strength;
\cite{sezgin1,sezgin2} constructed (1,0) superconformal models with tensor gauge
fields as well as Yang-Mills gauge fields, which are further analyzed in 
\cite{ana1,ana2}.

\section{Non-Abelian $SU(N_5)$ Self-Dual String Solution}

\subsection{Non-abelian theory of multiple M5-branes}

In \cite{CK}, an action for a non-abelian chiral 2-form in 6-dimensions was 
constructed as a generalization of the abelian theory of Perry-Schwarz \cite{PS}. 
As in Perry-Schwarz,
the self-dual tensor gauge field is represented 
by a $5 \times 5$ antisymmetric field $B_{\m\n}$, 
$\m, \n = 0,..,4$ with $B_{\m 5} =0$ and so 
manifest 6d Lorentz symmetry is given up. Presumably there is a covariant 
construction generalizing that of PST \cite{pst} where our theory will be obtained
after a certain gauge fixing.  

For $N_5$ coincident M5-branes, the self-duality equation of motion 
of the theory reads \cite{CK}
\be\label{sd-na}
\Ht_{\m\n} = \pa_5 B_{\m\n},
\ee
where the gauge field $A_\m$ is 
constrained by \eq{FB1}
and carries  no propagating degrees of freedom.
Here
\be
H_{\m\n\r}= D_{[\m}B_{\n\r]} = \pa_{[\m}B_{\n\r]}+[A_{[\m},B_{\n\r]}],
\ee
\be
\tilde{H}_{\m\n} = -\ove{6}\e_{\m\n\r\s\t}H^{\r\s\t},\qquad \e_{01234}=-1,
\ee
\be
F_{\m\n} = \pa_{[\m} A_{\n]} + [A_\m,A_\n]
\ee 
and are in the adjoint representation of $SU(N_5)$.
$c$ is a constant in the theory, which is fixed later by the
charge quantization condition of the self-dual string solution \cite{CKV}. 
Evidences have been given in \cite{CK,CKV} that this provides a 
description of the gauge sector of coincident M5 branes. Moreover, 
by combining knowledge of conformal symmetry and R-symmetry,
we have argued  that the 1/2 BPS equation in the case when 
only one scalar field is turned on takes the form
\be \label{bps}
H_{ijk}=\e_{ijk}\pa_5\f,\qquad
H_{ij5}=-\e_{ijk}D_k\f.
\ee 
This generalizes the 1/2 BPS monopole equation in 4-dimensional
non-abelian gauge theory.  Our convention for the Lie algebra are:
$[T^a, T^b] = i f^{abc} T^c$, 
$F_{\m\n}= i F_{\m\n}^a T^a$, $A_\m = i A_\m^a T^a$ and
$F_{\m\n}^a = \del_\m A_\n^a - \del_\n A_\m^a  - f^{abc} A^b_\m A^c_\n$.
In the abelian limit, the equations \eq{bps} reduce to that of \cite{HLW}.

\subsection{Generalized non-abelian Wu-Yang monopole}

Our construction of the self-dual string solution in the $SU(N_5)$ theory
will be based on a 
generalization of the $SU(2)$ charge one 
Wu-Yang monopole 
solution to one with arbitrary charge $n$. Let us start with
a review of the generalized Wu-Yang monopole.

Consider an $SU(2)$ gauge theory, and for reasons which will become clear later, 
let us denote the Lie algebra generators by  $\a^a$ ($a=1,2,3$) with
\be
[\a^a , \a^b] = i\e^{abc}\a^c.
\ee
The generalized Wu-Yang monopole is given by the following 
field configuration \cite{shnir}
\be \label{gen-A}
A_k = i A_k^a \a^a 
= -\frac{i}{2r} (\tn_\vphi \hat{\th}_k - n\tn_\th \hat{\vphi}_k), 
\ee
i.e. 
\be
A_\th = -\frac{i}{2r} \tn_\vphi,\quad 
A_\vphi = \frac{in}{2r} \tn_\th.
\ee
Here 
\bea
\hat{\th}_kdx^k &:=& \cos\th \cos\vphi dx^1 + \cos\th \sin\vphi dx^2 
-\sin\th dx^3,\nn\\
\hat{\vphi}_k dx^k &:=& -\sin\vphi dx^1 +\cos\vphi dx^2,
\eea
\be
x^1=r\sin\th\cos\vphi,\qquad
x^2=r\sin\th\sin\vphi,\qquad
x^3=r\cos\th,
\ee
\bea
\rn &:=&  (\sin\th \cos(n\vphi) , \sin\th \sin(n\vphi) , 
\cos\th),  \nn\\
\thn &:=& (\cos\th \cos(n\vphi) , \cos\th \sin(n\vphi) , -\sin\th) = 
\frac{\del\rn}{\del\th}, \nn\\
\vphin&:=& (-\sin(n\vphi) , \cos(n\vphi) , 0) = 
\frac{1}{n \sin\th} \frac{\del\rn}{\del\vphi}
\eea
and 
\be
\tn_r = \rn \cdot \vec{\a}, \quad 
\tn_\th = \thn \cdot \vec{\a}, \quad 
\tn_\vphi = \vphin \cdot \vec{\a}, 
\ee
The Wu-Yang monopole configuration considered in \cite{CKV} 
is given by the case $n=1$. 

It is useful to note that
\be
A_k = U \bar{A}_k U^{-1} + U\del_k U^{-1}, 
\ee
with
\be
U = e^{-i \vec{\a} \cdot \vphin}
\ee
and
\be
\bar{A}_k = \frac{n}{r} \frac{1-\cos\th}{\sin\th}\hat{\vphi}_k \times i\a^3, 
\ee
As a result, 
\be
F_{ij} = U \bar{F}_{ij} U^{-1}, 
\ee
where 
\be
\bar{F}_{ij} = \del_{i}\bar{A}_{j} - \del_j \bar{A}_i 
= n\e_{ijk} \frac{x^k}{r^3}  \times i\a_3.  
\ee
Since
\be
U \a^3 U^{-1} = \rn_a \a^a,
\ee
we obtain
\be \label{gen-F}
F_{ij} =  n\e_{ijk} \frac{x^k}{r^3} i \rn_a \a^a.
\ee
Therefore \eq{gen-A}, \eq{gen-F} provide 
a generalization of the Wu-Yang monopole solution to ``charge'' $n$.

We also note that
\be \label{Dr}
D_i \lrbrk{ \rn_a \a^a } = 0
\ee
holds for the generalized Wu-Yang monopole. This is
a generalization of $D_i ( x^a \a^a/r ) = 0$ for the charge one case.

\subsection{Non-abelian self-dual string and the distance-radius relation}

In this subsection we  
construct a self-dual string solution where the tensor gauge field
is embedded in an $SU(2)$ sub-algebra of $SU(N_5)$.
Let us denote the generators of the $SU(2)$ factor by $\a^a$: 
\be
[\a^a , \a^b] = i\e^{abc}\a^c. 
\ee
As an $N_5\times N_5$ irreducible 
representation $\cR$ of $SU(2)$,
the Casimir operator is given by
\be
\a_1^2 + \a_2^2 + \a_3^2 = \ove{4}(N_5^2 - 1) := R^2.
\ee
Therefore
\be
\tr(\a^a\a^b) = \mu^2 \d^{ab},
\ee
where
\be
\mu^2 := \frac{1}{12} N_5 (N_5^2 -1).
\ee

Consider an ansatz of the field strength
\be \label{H-an}
H_{\m\n\l} = H_{\m\n\l}^\PS \, i \rn_a  \a^a,
\ee
where $H^{(PS)}_{\m\n\l}$ is given by the abelian 
Perry-Schwarz solution \cite{PS} 
\be \label{H2-ab}
H_{045}^{(PS)}=\frac{\b x^5}{\r^4}, \qquad
H_{ij5}^{(PS)}=-\frac{\e_{ijk}\b x^k}{\r^4}
\ee
with  $\b$ being a constant to be fixed.
Obviously the field strength is self-dual. 
Since $\rn_a$ is independent of $x_5$, we can immediately 
integrate $H_{045}$, $H_{ij5}$ over $x_5$ and obtain
\be \label{eq:ansatz}
B_{\m\n} = B_{\m\n}^\PS \, i \rn_a \a^a,\qquad 
\m\n = \mbox{$ij$ or $04$},
\ee
where the Perry-Schwarz 2-form potential is: 
\be \label{soln-B1-ab}
B_{ij}^{(PS)}=-\hlf\frac{\b\e_{ijk}x_k}{r^3}\lrbrk{\frac{x^5r}{\r^2}
+\tan^{-1}\lrbrk{\frac{x^5}{r}}}, \qquad
B_{04}^{(PS)}=-\frac{\b}{2\r^2}.
\ee

One still need to check that this $B$-field reproduces correctly the other 
components $H_{ijk}$. 
To check this, we note that the constraint \eq{FB1}  gives 
\be
F_{ij} = -c\b \frac{\p}{2} \e_{ijm}\frac{x^m}{r} i \rn_a  \a^a  \qquad
F_{04} = 0. 
\ee
Therefore if we take
\be\label{bc-cond}
\b = -\frac{2 n }{ c \pi} , 
\ee
for an integer $n$, 
then  the auxiliary gauge field is given by the generalized Wu-Yang monopole.
As a result of \eq{Dr}, we have
\be
D_{[\l}B_{\m\n]} = \del_{[\l}B_{\m\n]}^\PS \, i \rn_a \a^a  
\ee
and the result agrees with \eq{H-an}. 

To obtain a self-dual string solution, we observe that 
the BPS equation \eq{bps} can be solved with
\be \label{sc}
\phi = - \lrbrk{ u + \frac{\b}{2\r^2} } i \rn_a \a^a.
\ee
This leaves an unbroken $U(1)$ generated by 
\be
\hat{\phi} := \phi/|\phi| =  \mp i \rn_a \a^a /\mu
\ee
at $\r \to \infty$. Here $|\phi|^2 := \tr(\phi^2)$ and
the $-(+)$ sign is chosen for $u>0 (u<0)$. Without loss of generality, we take
$u>0$ below. 
The asymptotic $U(1)$ fields are identified by a projection
\be
H_{\m\n\l}^\uo := \tr(H_{\m\n\l}\hat{\phi}), \qquad
B_{\m\n}^\uo := \tr(B_{\m\n} \hat{\phi}), 
\ee
then 
\be
H_{\m\n\l}^\uo 
= \mu H_{\m\n\l}^\PS, \qquad
B_{\m\n}^\uo 
=  \mu B_{\m\n}^\PS. 
\ee
The $U(1)$ magnetic and electric charges (per unit length) are defined by 
\be
2\pi^2 P = \int_{S^3} H^\uo , \quad 2 \pi^2 Q = \int_{S^3} *H^\uo.
\ee
This gives the $U(1)$ charges, 
\be \label{PQ}
P = Q = \mu \b
\ee
Note that our normalization for the charges 
differs from that of \cite{PS} by a factor of 
$2\pi^2$ of the volume of unit $S^3$. Our definition is consistent with the
Gauss law of the form: $ \nabla \cdot \vec{B} = 2\pi^2 P \delta^{(4)}(x)$, 
$ \nabla \cdot \vec{E} = 2 \pi^2 Q \delta^{(4)}(x)$ 
for a string lying in the $x^1$ direction, and $B_p := \e_{pqrs} H^{qrs}/3!$,
$E_p :=  H_{01p}$, $(p =2,3,4,5)$.
With this normalization of charges, the 
charge quantization condition for dyonic strings in an abelian theory of 
2-form in six-dimensions 
reads \cite{deser1,deser2}
\be \label{deser}
PQ' + QP' = 2\pi \frac{\bb{Z}}{(2\p^2)^2}. 
\ee

For us, the charge quantization condition is modified 
due to the existence of a non-trivial center in the residual 
gauge group of the non-abelian theory. In fact, in  
a non-abelian Yang-Mills gauge theory with an 
unbroken gauge group of the form 
\be
H = U(1) \times K,
\ee
where the $U(1)$ factor allows one to define the electric and
magnetic charges, and $K$ is any residual gauge group, 
Corrigan and Olive have shown \cite{corrigan} that 
the charge quantization takes the form
\footnote{The normalization of the magnetic charge was taken as
$\nabla \cdot \vec{B} = 4 \pi g$ 
in \cite{corrigan}. 
}
\be
e^{4\pi i q g} =k,
\ee
where $k$ is an element in $C(K)$, the center of $K$. 
For example, if $K= SU(N)$, then $C(K) =Z_N$, 
\be
k = e^{2\pi i n/N} \id_N, \quad n = \mbox{integers}
\ee
and the charge quantization condition for the monopoles reads
\be
q g = \frac{n}{2N}.
\ee
For us, since 
the symmetry is broken down by the scalar field as
$SU(N_5) \to U(1) \times SU(N_5-1)$ and since 
(2,0)  supersymmetry demands that all fields in theory 
to be in the adjoint representation, this means  the center of the residual gauge
symmetry is given by $Z_{N_5-1}$. 
The same argument 
as Corrigan and Olive then 
gives
\be \label{cq1}
PQ' + QP' = \frac{2 \pi}{N_5-1} \frac{\bb{Z}}{(2\p^2)^2}. 
\ee

We can now use the charge quantization condition \eq{cq1} to fix the value of $\b$.
Let us consider the situation where the self-dual string configuration arises as 
the intersection of a number $N_2$ of coincident M2-branes ending 
perpendicularly on our system of M5-branes. 
In this case, the charges $P, Q$ 
should be proportional to $N_2$.
Substituting \eq{PQ}, we obtain
\be
P = Q = \mu \b= \frac{N_2}{\sqrt{N_5-1} } Q_0, 
\ee
where $Q_0 = \sqrt{\pi}/(2\p^2)$ is the minimal unit of charge in the abelian theory
\cite{PS,HLW}. 

From the field theory point of view, the geometrical shape of the
M2-branes spike is described by the scalar field 
$X= 4 \phi$.
Here the normalization factor of 4 was obtained 
from the analysis of \cite{HLW}
where the geometrical relations between target space coordinates and worldvolume
scalar field is the clearest in the superembedding formalism. 
It is convenient to introduce the root-mean-square distance (setting $l_p =1$)
\be
D: = \sqrt{\frac{1}{N_5} |\tr( 4 \phi)^2 |}
\ee
as a measure of the transverse distance of the M2-branes spike
from the system
of M5-branes.  The cross section of the 
M2-branes spike is an $S^3$ and the radius $\rho$ is governed by the 
transverse distance-radius relation
\be \label{Dx}
D =  D_0
+ \frac{2N_2 Q_0}{\sqrt{N_5 (N_5-1)} \r^2}
\ee
where 
\be
D_0 := \frac{4 u\mu}{\sqrt{N_5}}
\ee
is a constant.

In addition to the worldvolume description, 
the system of intersecting M2-M5 branes also admits a supergravity description
from which one can extract the transverse distance-radius relation 
and compare with our field theory result.  However, 
the supergravity solution, beyond the brane probe approximation, 
for a system of M2-branes
intersecting on a  system of separated M5-branes where
$(N_5-1)$ of them are in coincidence and another single M5-brane is separated at a
finite distance away from the main group
is not known. 
The closest 
system which admits a supergravity solution is the system
of $N_2$ M2-branes
intersecting a system of $N_5$ coincident M5-branes \cite{siampos}. In this paper
the technique of blackfold is applied and the 
transverse distance-radius relation
\be \label{Dx1}
D =  \frac{2 \pi N_2}{N_5} \,\frac{1}{\rho^2} 
\ee
is obtained. 
At distance $D$ large enough compared with the separation so that one cannot 
resolve the  details of the separation, one can expect the supergravity 
solution for our system can be approximated by the supergravity 
solution of this system. 
In this case,
one can ignore the first term in our 
field theory result \eq{Dx} and our 
transverse distance-radius relation 
\be \label{Dx2}
D =  \frac{2N_2 Q_0}{N_5 \r^2}
\ee
agrees precisely 
with that of supergravity on their  $N_2$ and $N_5$ dependence. Note that, 
however, \eq{Dx1} and \eq{Dx2} differs by an overall scale
factor of $2\p^2\sqrt{\pi}$. 
This is  presumably due to a different unit is employed in supergravity.

We remark that the field theory description and the supergravity description are
good  only in their respective regime of validity. To confirm the 
validity of the agreement we found above, one need to check that there 
indeed exists an overlapping regime where one can trust both descriptions 
and hence compare the results 
sensibly. To check this, we note that
our  description of the M2-branes spike 
as a worldvolume soliton of the M5-branes is good provided that higher
derivative corrections to our non-abelian action is
small: $l_p|\pa^2\Phi| \ll |\pa\Phi|$.
This translates to
\be\label{lowlim}
\r \gg l_p.
\ee
On the other hand, the validity of the blackfold
description \cite{siampos} was discussed in \cite{siampos2}.
It was found there that for zero angular velocity which is our case, one needs
\be \label{lowlim2}
\r \gg \text{max}(l_p,\r_c),
\ee
where
$ \r_c := N_5^{1/3}( 1+\sqrt{1+64 N_2^2/N_5^4}) ^{1/6}l_p.$
Therefore in the region \eq{lowlim2}, both the supergravity description and the 
M5-branes worldvolume description are valid. Now in order for the field theory 
result \eq{Dx} to reduce to the form \eq{Dx1}, it is required that
\be \label{co2}
\rho \ll \lrbrk{\frac{3}{4\p^3}}^{1/4}\lrbrk{\frac{ N_2}{u N_5^2}}^{1/2}.  
\ee 
Thus, 
by arranging the 
parameters $N_2$, $N_5$ and $u$
(for example a scaling limit involving
large $N_2$, $N_5$ and small $u l_p^2$),
a non-empty region of $\r$ 
satisfying both \eq{lowlim2} and \eq{co2} can always be achieved,
and  so the agreement we found is justified.

Summarizing, we can take the  constant $c$ of the action to be
$c= - 2\mu \sqrt{N_5 -1} /(Q_0\p)$.
Then
\eq{eq:ansatz}, \eq{sc} provide a self-dual string solution to the theory
if we take
\be 
\b  = \frac{N_2 Q_0}{ \mu \sqrt{N_5-1}}.
\ee

\section{Discussions}

In this paper, we have constructed a self-dual string solution in the $SU(N_5)$
non-abelian theory of five-branes.  Our solution carries an arbitrary 
$N_2$ unit of self-dual charge and  obtains its charge through the generalized
non-abelian Wu-Yang monopole configuration carried by the auxiliary one-form 
gauge field. We have also shown that the 
radius-transverse distance relation describing the M2-branes spike
agrees precisely
with the supergravity description of the intersecting M2-M5 branes 
system. Our results in this paper therefore provide further evidence that the 
non-abelian theory \cite{CK} does give a description for a 
system of multiple M5-branes. 

Our self-dual string solution was obtained for a special symmetry breaking where
there is a residual $SU(N_5-1)$ gauge symmetry. For a more general configurations
of the Higgs field, the residual symmetry at infinity could be 
smaller and the self-dual string would
be characterized by more number of charges. The discussion is similar to that of
non-abelian
monopole \cite{manton}. It will be interesting to construct these other kinds of 
self-dual strings and understand their dual description in 
supergravity and M-theory.

Our results were derived by solving the self-dual string equation \eq{bps}. 
As argued in \cite{CKV}, 
this could be obtained from the 1/2 BPS condition of a supersymmetry 
transformation law of the form
\be\label{tf}
\d \psi = (\G^M \G^I D_M \phi^I + \frac{1}{3!2} \G^{MNP} H_{MNP}) \e.  
\ee
Our results thus provide support that 
\eq{tf} is indeed the correct supersymmetry transformation law for the (2,0) 
supersymmetric 
theory. This additional information should be helpful for tackling the so far 
mysterious nature \cite{CK} of the (2,0) supersymmetry. 
It would also be interesting to consider other BPS solutions and check them against
the supergravity description.

With the solution at hand, one may perform a small fluctuation analysis 
of the solution and use it 
to 
learn more about the dynamics of non-abelian self-dual string. 
It would also be  interesting to include couplings to a background  $C$-field.
In \cite{CG,qg}, the quantum Nambu geometry has been obtained
as the quantum geometry for M5-branes in a large constant background $C$-field.
One should be able to construct the star-product for this geometry and use it
to derive the ``Seiberg-Witten'' map for the  non-abelian M5-branes theory 
in a background $C$-field.

Finally we note that 
our field theory description is not good near the spike region ($\r\simeq 0$).
Apparently,
unlike the non-abelian BPS monopole, the non-abelian interaction here is not 
sufficient to remove the spike singularity. Nevertheless one can 
expect it to be smoothen out only in the complete
description of M-theory with all higher derivative corrections included \cite{PS}.

\section*{Acknowledgements}

It is a pleasure to thank 
Neil Lambert, 
Christian Saemann, 
Gary Shiu, Douglas Smith, Henry Tye,  Martin Wolf
as well as participants of the IAS Durham workshop
``Symmetry and Geometry of Branes in String/M Theory''
for useful discussions. We also thank Sheng-Lan Ko for discussions and
comments on the manuscript.
CSC is supported in part 
by the National Science Council, Taiwan.
PV is supported by a Durham Doctoral Studentship 
and by a DPST Scholarship from the Royal Thai Government.

\end{document}